\documentclass[aps,prl,showpacs,twocolumn]{revtex4}
\usepackage[dvips]{graphicx}
\usepackage{epsfig}
\usepackage{amsmath}
\usepackage{amsfonts}
\usepackage{amssymb}
\usepackage{wasysym}

\def\be{\begin{equation}}
\def\ee{\end{equation}}
\def\bfi{\begin{figure}}
\def\efi{\end{figure}}
\def\bea{\begin{eqnarray}}
\def\eea{\end{eqnarray}}

\begin{document}


\title{The balance between excitation and inhibition controls 
the temporal organization of neuronal avalanches}

\author{F. Lombardi$^{1}$, H. J. Herrmann$^{1,2}$,
C. Perrone-Capano $^{3}$, D. Plenz $^{4}$ and  L. de Arcangelis$^{5}$ }

\affiliation{
$^{1}$  Institute Computational Physics for Engineering Materials, 
ETH, Z\"urich, CH\\
$^{2}$ Departamento de F\'{\i}sica, Universidade Federal do Cear\'a,
60451-970 Fortaleza, Cear\'a, Brazil\\
$^{3}$ Biological Sciences Dept., Univ. of Naples Federico II and IGB-CNR, 
Napoli, Italy\\
$^{4}$ Section on Critical Brain Dynamics, NIH, Bethesda, Maryland
20892, USA\\
$^{5}$ Dept. of Information Engineering,
Second University of Naples, Aversa (CE), Italy
}

\begin{abstract}
Neuronal avalanches, measured in vitro and in vivo, exhibit a robust critical
behaviour. Their temporal organization hides
the presence of correlations. 
Here we present experimental measurements 
of the  waiting time distribution between successive avalanches 
in the rat cortex
in vitro. This exhibits a  non-monotonic behaviour, 
not usually found in other
natural processes. Numerical simulations provide evidence that this 
behaviour is a consequence of
the alternation between states of high and low activity, named up and down
states, leading to a balance between excitation and inhibition controlled by
a single parameter. During these periods both 
the single neuron state and the network excitability level, keeping memory of past
activity, are tuned by homeostatic mechanisms. 

\end{abstract}


\maketitle

Spontaneous neuronal activity can exhibit slow oscillations between 
bursty periods, or up-states, followed by substantially quiet periods.
Bursts can last from a few to several hundreds of
milliseconds and, if analysed at a finer temporal scale, have often 
shown a complex
structure in terms of neuronal avalanches.
{\it In vitro} experiments record avalanche activity\cite{beg1,beg2}
from mature organotypic cultures of rat somatosensory
cortex where they spontaneously emerge in superficial layers.
The size and duration of
neuronal avalanches follow power law distributions with stable exponents,
which is a typical feature of a system in a
critical state, where large fluctuations are present and the response does
not have a characteristic size. The same critical behaviour 
has been measured  {\it in vivo} from rat cortical layers during early
post-natal development\cite{pnas}, from the cortex of
awake adult rhesus monkeys\cite{pnas2},
using microelectrode array recordings, as well as for
dissociated neurons from rat hippocampus\cite{maz,pas} or leech ganglia\cite{maz}.
{\it In vitro},  quiet periods measured between bursts, also 
called down-states, can last up to several seconds. The emergence of these 
down-states 
can be attributed to various mechanisms: a decrease in the 
neurotransmitter 
released, either due to the exhaustion of available synaptic 
vesicles or to the increase of a factor inhibiting the 
release\cite{stal} such as the nucleoside adenosine\cite{thom},
 the blockade of receptor channels by the presence 
of external magnesium\cite{mae}, or else spike 
adaptation\cite{mcco}.
A down-state is then characterized by a {\it disfacilitation}, i.e. 
absence of 
synaptic activity, of a large number of neurons causing long-lasting returns 
to resting potentials\cite{timo}.
Recently, it was shown analytically and numerically that 
critical behaviour\cite{lev12} characterizes up-states, whereas 
down-states are subcritical\cite{nieb}.

Whereas action potentials are rare during down-states, small amplitude 
depolarizing potentials, reminiscent of miniature potentials from spontaneous
synaptic release,   
occur at higher frequencies. The non-linear 
amplification of small amplitude signals contributes to the generation of larger 
depolarizing events bringing the system back into the up-state, as observed in
cortical slabs\cite{timo2}, dissociated cultures\cite{mar} and slice cultures\cite{plae}.  
The analysis of the amount of time striatal spiny neurons\cite{plki,jw} and
cortical pyramidal neurons\cite{cunn} spend at each value of the membrane potential
shows that both cell types toggle between two preferred values\cite{has}:
a very
negative one in the down state, and a more positive, depolarized one, in the
up-state.  The up-state being just a few millivolts from the action potential
threshold, suggests that during the up-state neurons respond faster and more
selectively to synaptic inputs.
For cortical neurons the up-state would be a metastable state, i.e. the membrane 
potential would soon decay down to the resting potential value, 
if network mechanisms 
would not sustain the activity. The up-state has therefore network, rather than 
cellular, properties. 

 
Here we focus on the temporal organization of neuronal avalanches
both in organotypic cultures and neuronal networks simulations.
Each avalanche $i$ is characterized by  its starting and ending 
times, $t^i_i$ and $t^f_i$. The 
temporal organization is analysed by evaluating the distribution of
waiting times $\Delta t_i =t^i_{i+1}-t^f_i$. 
This is a fundamental property of  stochastic 
processes, widely investigated for  natural phenomena
and able to discriminate between a simple Poisson and a 
correlated process. Indeed, in the first case the distribution is
exponential, whereas it exhibits a more complex
behaviour with a power law regime if correlations are present.
For a wide variety of phenomena, e.g. earthquakes and solar flares \cite{dea}, 
human dynamics \cite{bara}, biological systems \cite{bru},
etc., this distribution always shows a monotonic behaviour.
Recent results on freely
behaving rats provide a lognormal size distribution and a monotonic waiting
time distribution uniquely controlled by
the average occurrence rate\cite{rib}. Anaesthetized rats, conversely,
exhibit a more heavy-tailed size distribution and
no universal scaling for the waiting time distribution.
Here we show that the waiting time distribution for neuronal avalanches {\it in
vitro} has an unusual non monotonic behavior. Numerical simulations on neuronal
networks suggest that this is controlled by the slow 
alternation of up and down states, which determines both the 
network and the single neuron behavior.

\begin{figure}
\includegraphics[width=8cm]{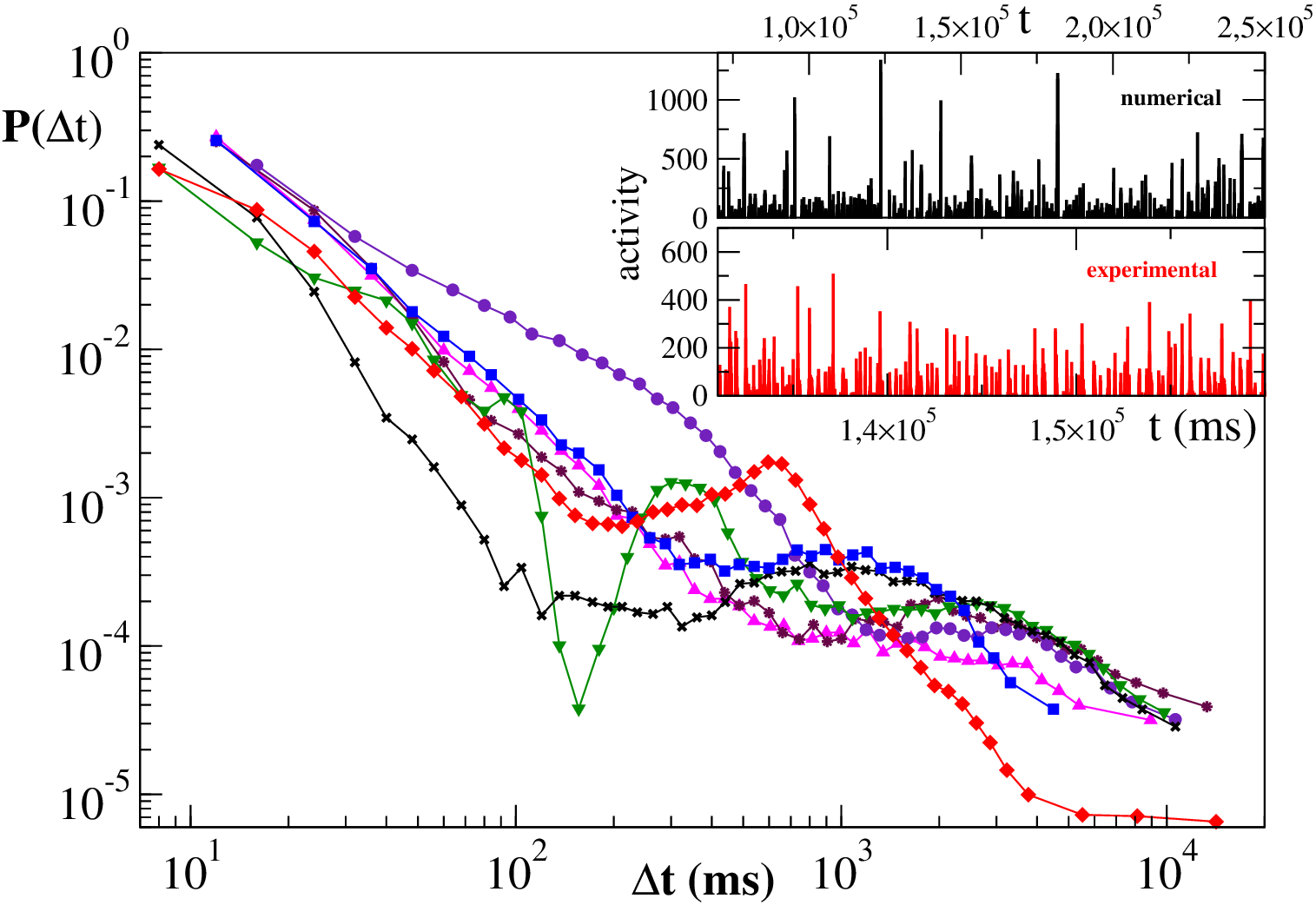}
    \caption{(Color online)
The distribution of waiting times for seven different slices of rat cortex. 
All curves show an initial power law regime between 10 and
about 200ms, with an average
exponent $2.15 \pm 0.32$. For $\Delta t>200ms$ curves can become
quite different with the common characteristics of a local minimum located
at  $200ms<\Delta t_{\rm min}<1s$, followed by a more or less pronounced
maximum at $\Delta t\simeq 1-2s$. In the insets two temporal
sequences of neuronal activity for numerical (sum of potential variations) 
and experimental ($\mu V$) data.
}
\label{fig1}
\end{figure}

Experiments were performed on coronal slices from rat dorsolateral cortex 
(postnatal day 0-2; 350 $\mu m$ thick)
attached to a poly-D-lysine coated 60-microelectrode array (MEA;
Multichannelsystems,
Germany) and grown at 35.5 $^o$C in normal atmosphere and standard culture medium
without antibiotics for 4-6 weeks before recording.
Avalanche activity was measured from cortex-striatum-substantia nigra
triple cultures or single
cortex cultures as reported previously\cite{beg1}. Spontaneous avalanche
activity is recorded
outside the incubator in standard artificial cerebrospinal fluid (ACSF; laminar
flow of ~1 ml/min) under stationary conditions for up to 10 hrs.
 The spontaneous local field potential (LFP) is sampled continuously at 1
kHz at each electrode and low-pass filtered at 50 Hz. Negative deflections in
the LFP (nLFP) are detected by crossing a noise threshold of -3 SD 
(SD stands for standard deviation $\sim 3-5 \mu V$) followed
by negative peak detection within 20 ms.
nLFP times and nLFP amplitudes are measured.
Neuronal avalanches are defined as spatio-temporal clusters of nLFPs on
the MEA\cite{plch}. 
A neuronal avalanche consists of a consecutive series of
time bins of width $\delta t$ that contain at least one nLFP on any of the
electrodes. Each avalanche is preceded and ended by at least one time bin with 
no activity. The waiting time $\Delta t$ is simply given by the number of empty
bins between two successive avalanches times $\delta t$. Without loss of
generality, the present analysis is done with $\delta t$ 
estimated for each culture as the average inter nLFP interval on the array
and ranged between 3-6 ms for all cultures.

In Fig.1 we show the waiting time distribution for different cultures of
rat cortex slices.
The curves exhibit a complex non-monotonic behaviour 
with common features: an initial power law regime and a local minimum 
followed by a more or less pronounced maximum. 
The presence of a power law
implies that avalanche occurrence is not a pure Poisson process, namely
successive avalanches are temporally correlated \cite{utsu}.
Moreover, the non-monotonic behaviour is not usually observed in natural
phenomena.  
In order to investigate the origin of this behaviour, we simulate avalanche 
activity
by a neuronal network model\cite{br1,br2,br3}, which  is
able to reproduce the scaling properties of neuronal avalanches. 
Here we question whether and how the complex temporal organization of avalanches
can  be caused by the  slow alternation between up-states and down-states.
The basic idea is that after a large avalanche the involved neurons become 
hyperpolarized and the system goes into a down-state. 
Conversely, after a small avalanche active neurons remain depolarized, the 
system stays in an up-state. 

We consider $N$ neurons at random positions, characterized by their potential
$v_i$. Neurons are connected by a classical scale-free network \cite{chia2}, where
a neuron $i$ has an out-going connectivity degree $k_{out_i}$. 
Once the network of output connections is established, we identify the
resulting degree of in-connections, $k_{in}$ for each neuron.
To each synaptic connection we assign an
initial random strength $g_{ij}$, with $g_{ij}\neq g_{ji}$, and to each neuron
an excitatory or inhibitory character, with 10\%
inhibitory synapses. Whenever at a given time  the
value of the potential at a site $i$ is above a certain threshold,
$v_i \geq v_{\rm max}$, the neuron sends action potentials
which arrive to each of the $k_{out_i}$ pre-synaptic buttons. 
As a consequence, the total charge  entering  the connected neurons 
is $q_i\propto v_i k_{out_i}$, as in a firing rate based charge distribution.
Each neuron receives charge in proportion to the synaptic strength 
$g_{ij}$,
$v_j(t+1)=v_j(t)\pm \frac{q_i(t)}{k_{in_j}}\frac{g_{ij}(t)}{\sum_k g_{ik}(t)}$,
where the sum is on all out-going
connections of $i$. Here the membrane potential variation is obtained
by dividing the received charge by
the surface of the soma of the post-synaptic neuron, proportional
to the number of  in-going terminals $k_{in_j}$. The plus or minus sign
is for excitatory or inhibitory $g_{ij}$, respectively.
After firing, a neuron is set in a refractory state lasting $1$ time
step (about 10 ms),
during which it is unable to  receive or transmit any charge.
At the end of an avalanche, we implement Hebbian plasticity rules:
The strength of the used connections between active
neurons  is increased proportionally to
the activity of the synapse \cite{coo}, namely the
membrane potential variation of the post-synaptic neuron, 
$g_{ij}(t+1) =g_{ij}(t) +(v_j(t+1)-v_j(t))/v_{\rm max}$.
Conversely, the strength of all inactive synapses is
reduced by the average strength increase per bond,
$\Delta g = \sum_{ij, t} \delta g_{ij} (t)/ N_b$,
where $N_b$ is the number of bonds. 
The presence of both strengthening and weakening rules implements a
homeostatic regulatory mechanism for synaptic strengths, which 
underlies the system's critical behaviour.
An external stimulus triggers further activity in the system:  
at the end of each avalanche the potential of a random neuron is increased by
a small amount until another neuron gets at threshold and starts an avalanche.
We implement the plasticity rules during a series of stimuli in order
to modify the synaptic strengths, initially random.
Previous studies have verified that the critical 
behaviour of avalanche distributions does not depend on parameter values
or network properties and that this model reproduces quantitatively
the background spectrum of measured EEG signals\cite{br1,br2,br3}.
The implementation of a scale free network of connections in the present
study is motivated by numerical convenience in terms of cpu time.

In order to 
implement the alternation between up and down-states,
at the end of each avalanche we measure 
its size in terms of the sum of depolarizations $\delta v_i$ of 
all active neurons, $s_{\Delta v}=\sum \delta v_i$.
If the last avalanche is larger than a threshold,
$s_{\Delta v}>s_{\Delta v}^{min}$, the system transitions into a down-state and
neurons active in the last avalanche become hyperpolarized
proportionally to their previous activity, namely 
we reset
\be
v_i=v_i - h \delta v_i
\ee
where $h>0$. 
This rule introduces a short range memory at the level of a single neuron
and models 
the local inhibition experienced by a neuron, due to
spike adaptation, adenosine accumulation,
synaptic vesicle depletion, etc. 

Conversely, if the avalanche just ended has a size
$s_{\Delta v}\le s_{\Delta v}^{min}$, the system either will remain,
or will transition into an up-state.
All neurons firing in the previous avalanche 
are set to the depolarized value
\be
v_i=v_{\rm max}(1-s_{\Delta v}/s_{\Delta v}^{min})
\ee
The neuron  potential depends  on the response of
the whole network via 
$s_{\Delta v}$, in agreement with measurements of
the neuronal membrane potential  which remains
close to the firing threshold in the up-state. 
$s_{\Delta v}^{min}$
controls the extension of the up-state and therefore
the level of excitability of the system.
The high activity in the up-state must be sustained by collective effects in
the network, otherwise the depolarized potentials would soon decay to zero, and
therefore the random stimulation in the up-state has an amplitude that depends on past
activity.
Eqs. (1) and (2) 
each depend on a single parameter, $h$ and $s_{\Delta v}^{min}$, which
introduce a memory effect at the level of single neuron
activity and the entire system, respectively. 
In order to reproduce the behavior observed experimentally, the
parameters  $s_{\Delta v}^{min}$ and $h$ are controlled separately.
Our simulations will show that the ratio $R=h/s_{\Delta v}^{min}$ 
is the only relevant
quantity controlling the temporal organization of avalanches. 

\begin{figure}
\includegraphics*[width=6.5cm]{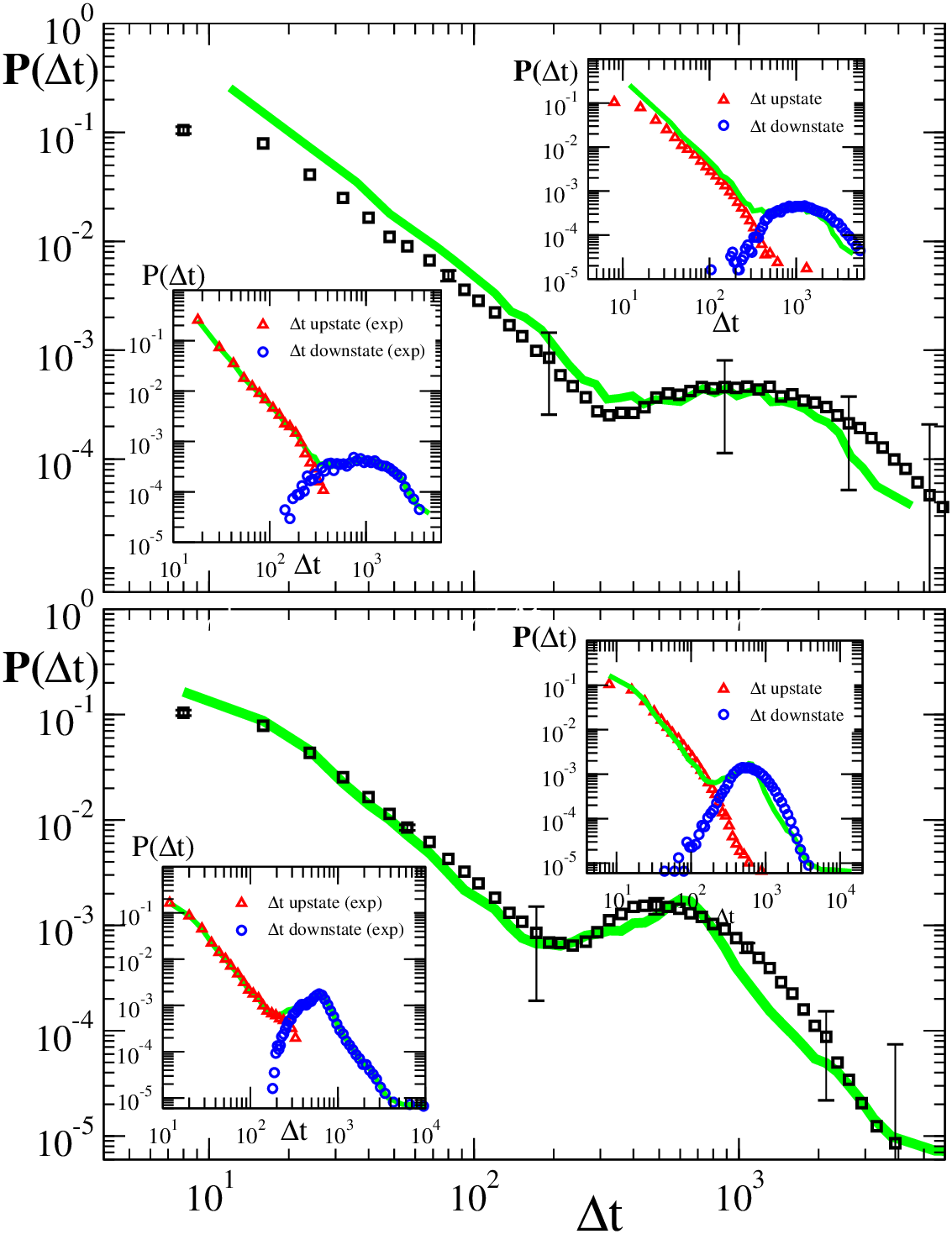}
    \caption{(Color online) Waiting time distributions measured experimentally
are compared with the average  numerical distributions for 
100 networks with $N=64000$ neurons.
Top: numerical curve ($s_{\Delta v}^{min}=140$ and $h=0.017$) 
fitting  the experimental
curve with blue squares in Fig.1;
Bottom: numerical curve ($s_{\Delta v}^{min}=110$ and $h=0.02$) fitting the
experimental curve with red diamonds in Fig.1. 
In the insets the waiting time distribution evaluated separately 
in the up and downstate for the numerical (upper insets) and the experimental
curves (lower insets).
For the numerical curves, statistical error bars  not shown are 
comparable to the symbol size.
     }
    \label{fig2}
\end{figure}

Numerical simulations show that the system indeed switches
between up and down states, with different
temporal durations (insets of Fig.1). 
The numerical waiting time distributions (Fig.2) exhibit
the non-monotonic behaviour of the experimental curves, where the position of 
the minimum
is controlled by the value of $s_{\Delta v}^{min}$ and the power law regime 
scales with the same exponent $\sim -2$ as experimental data.
The agreement between the numerical and the experimental distribution
is confirmed by the Kolmogorov-Smirnov test at a $p=0.05$ significance level.
Both distributions pass the statistical test with $p=0.99$ (bottom panel)
and $p=0.68$ (top panel). 
The different contribution from the two
states is reflected in the activity temporal scale (insets of Fig.2). 
The up-state generates
strongly clustered avalanches, originating the power law regime of the
waiting time distribution, 
whose extension depends on $s_{\Delta v}^{min}$. Large $\Delta t$ between 
avalanches generated in the upstate are observed with
a very small probability, which  increases with decreasing $h$.
Conversely, the waiting time distribution evaluated in the down-state 
has a bell-shaped behaviour centered at large intertimes which depends on
 $h$, i.e. for
a larger disfacilitation of the network the probability to observe intermediate
waiting times decreases in favour of long $\Delta t$.

The presence of the minimum and the height of the relative maximum are 
sample dependent (Fig.1) and for each sample 
the agreement between numerical and experimental data depends on the subtle 
balance between excitation and inhibition. For different samples, 
optimal agreement is realized when
the ratio $R=h/s_{\Delta v}^{min}\simeq 10^{-4}$.
For
instance, enhancing excitation, by increasing the threshold value
$s_{\Delta v}^{min}$, clearly produces a major shift in the data (Fig.3). 
Increasing inhibition, by increasing the
parameter $h$, generates the opposite effect, recovering the good agreement 
with experimental data. Interestingly, the avalanche
size and duration distributions also reproduce the experimental scaling
behaviour for the parameter values expressing the
balance between excitatory and inhibitory components. 
The abrupt transition between the up
and down-state, controlled by a threshold mechanism, generates
the minimum observed experimentally. 
Simulations of
up-states and down-states only in terms of different external drives,
without the single neuron state dependent behaviour (Eq.s 1-2), provide a 
monotonic waiting time distribution (inset Fig.3).

This complex non-monotonic behaviour, controlled
by the system balance level between excitation and inhibition expressed by the
parameter $R$, does not simply depend on the occurrence rate.
The different behaviour with respect to alive rats \cite{rib} 
could be attributed to
a larger separation in characteristic temporal scales between up and 
down-states.   Indeed, long lasting down-states in our case
originate waiting times one order of magnitude longer than for awake rats.  
Avalanches are temporally correlated in the up-state, whereas down-states are
long term recovery periods where memory of past activity is erased.
A detailed analysis of power spectra may shed further light on the 
temporal features of this alternation.
The good agreement with experimental data indicates that
the transition from an up-state to a down-state has a high degree of 
synchronization. Moreover it confirms that alternation between up and
down-states is the expression of a
homeostatic regulation which, during periods of high activity,    
is activated to control the excitability of the system and avoid pathological
behaviour.
The model suggests that the crucial feature of this temporal evolution
is the different single neuron behaviour in the two phases:
These collective effects must be supported by the single neuron behaviour, which
toggles between two preferential states, a depolarized one in the up-state and
a hyperpolarized one in the down-state. The model suggests that the depolarized
neuron state is a network effect: the avalanche activity itself determines how 
close to the 
firing threshold a neuron stays in the up-state. Conversely, the hyperpolarized
state is a form of temporal auto-correlation in the neuron activity.
The critical state realizes the correct
balance between excitation and inhibition via these self-regulating mechanisms.

\begin{figure}
\includegraphics*[width=7cm]{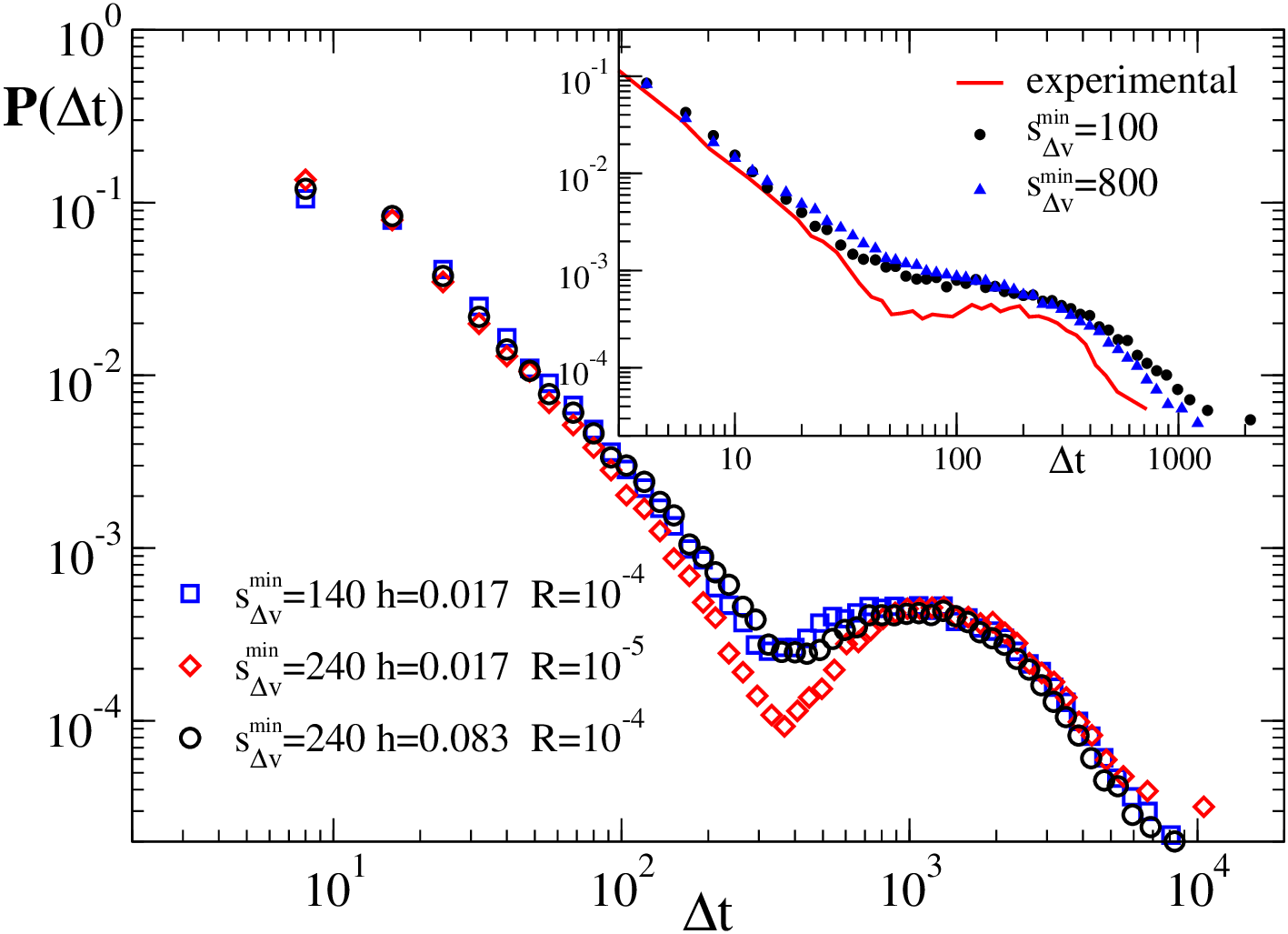}
    \caption{(Color online)
Waiting time distribution measured 
numerically for 100 networks of $N=16000$
neurons with different $s_{\Delta v}^{min}$ and $h$. 
The best agreement is obtained for $R\simeq 10^{-4}$.
In the inset waiting time distributions obtained with different stimulations
in the up and down-states and without the single neuron state behaviour ($h=0$).
     }
    \label{fig:fig3}
\end{figure}

{\it Acknowledgments.}
We thank the SNF for funding within project 205321-13874.
DP is supported by the Intramural Research Program of the NIMH, NIH.


\begin{thebibliography}{references}

\bibitem{beg1} J.M. Beggs, D. Plenz, J.  Neurosci. {\bf 23}, 
11167 (2003). 

\bibitem{beg2} J.M. Beggs, D. Plenz, J. Neurosci. {\bf 24}, 5216 (2004).


\bibitem{pnas} E.D. Gireesh, D. Plenz,  Proc. Nat. Acad. Sci. USA {\bf 105}, 
7576 (2008).

\bibitem{pnas2} T. Petermann et al,
Proc. Nat. Acad. Sci. USA {\bf 106}, 15921 (2009).

\bibitem{maz} A. Mazzoni et al,
PLoS ONE 2:e439 (2007).

\bibitem{pas} V. Pasquale et al,
Neuroscience {\bf 153}, 1354 (2008).


\bibitem{stal} K.J. Staley et al, 
Nature Neurosci.  {\bf 1}, 201 (1998).

\bibitem{thom} S.M. Thompson et al, 
J. of Physiol. {\bf 451}, 347 (1992).

\bibitem{mae} E. Maeda et al, 
J. Neurosci. {\bf 15}, 6835 (1995). 

\bibitem{mcco} M.V. Sanchez-Vives et al, 
J. Neurosci. {\bf 20}, 4286 (2000).


\bibitem{timo} I. Timofeev et al, 
Proc. Nat. Acad. Sci.  USA {\bf 98}, 1924 (2001).


\bibitem{lev12} A. Levina et al, 
Nature Phys. {\bf 3}, 857 (2007);  Phys. Rev. Lett. {\bf 102}, 118110, (2009).

\bibitem{nieb} D. Millman et al, 
Nature Phys. {\bf 6}, 801 (2010).

\bibitem{timo2} I. Timofeev et al,
Cer. Cortex {\bf 10}, 1185 (2000).

\bibitem{mar} D. Eytan, S. Marom, J. Neurosci. {\bf 26}, 8465 (2006).

\bibitem{plae} D. Plenz, A. Aertsen, Neuroscience {\bf 70}, 893 (1996).

\bibitem{plki} D. Plenz, S.T. Kitai, J. Neurosci. {\bf 18}, 266 (1998).

\bibitem{jw} E.A. Stern et al, 
Nature {\bf 394}, 475 (1998).

\bibitem{cunn} M.O. Cunningham et al, Proc. Nat. Acad. Sci. USA {\bf 103}, 5597 
(2006).

\bibitem{has} A. Hasenstaub et al, 
J. Neurosci. {\bf 27}, 9607 (2007).

\bibitem{dea} L. de Arcangelis et al, 
Phys. Rev. Lett. {\bf 96}, 051102 (2006). 

\bibitem{bara} A.L. Barabasi, Nature {\bf 435}, 207 (2005);
D. Rybski et al, Proc. Nat. Acad. Sci. U.S.A. {\bf106}, 12640 (2009). 

\bibitem{bru} J. Brujic et al, Nature Physics {\bf 269}, 282 (2006).

\bibitem{rib} T.L. Ribeiro et al,
PLoS ONE 5:e14129 (2010).


\bibitem{plch} D. Plenz, D.R. Chialvo,  arXiv:0912.5369.

\bibitem{utsu} T. Utsu, 
International Handbook of Earthquake and Engineering
Seismology, {\bf 81A}, 719 (2002).

\bibitem{br1} L. de Arcangelis et al, 
Phys. Rev. Lett. {\bf 96}, 028107 (2006).
 
\bibitem{br2} G.L. Pellegrini et al, 
Phys. Rev. E {\bf 76}, 016107 (2007).

\bibitem{br3} L. de Arcangelis, H.J. Herrmann, Proc. Natl. Acad. Sci. USA 
{\bf 107}, 3977 (2010).

\bibitem{chia2} V.M. Eguiluz et al,
Phys. Rev. Lett. {\bf 94}, 018102 (2005).


\bibitem{coo} S.J. Cooper, Neurosci. Biobehav. Rev. {\bf 28}, 851
(2005).







\end{thebibliography}
\end{document}